# Effect of Se doping in recently discovered layered $Bi_4O_4S_3$ Superconductor

Rajveer Jha and V.P.S. Awana*

CSIR-National Physical Laboratory (CSIR), Dr. K. S. Krishnan Road, New Delhi 110012, India

We report suppression of superconductivity in $Bi_4O_4S_3$ compound by Se doping at S site. Bulk polycrystalline samples are synthesized by solid state reaction route. The Rietveld refined of XRD data of all the studied samples show that the same are crystallized in tetragonal $I4/mmm$ space group with slightly increased c lattice constant with the Se doping. Superconductivity is observed at below 4.5 K in both dc magnetization and resistivity with temperature measurements for all $Bi_4O_4S_{3-x}Se_x$ (x = 0, 0.03, 0.09 and 0.15) samples, though the same is decreased slightly with increase in Se content. The upper critical field of the $Bi_4O_4S_3$ is estimated from resistivity under magnetic field [$\rho(T)H$] measurements of up to 2T (Tesla), and the same decreases for Se doped samples. The flux flow activation energies being obtained by fitting to the the Arrhenius equation are 18.60 meV for $Bi_4O_4S_3$, 13.69 meV for $Bi_4O_4S_{2.97}Se_{0.03}$ and 13.16 meV for $Bi_4O_4S_{2.85}Se_{0.15}$ samples in applied magnetic field of 200 Oe. In conclusion the S site Se substitution showed detrimental effect on superconductivity of $Bi_4O_4S_3$.

Introduction

The discovery of superconductivity with superconducting transition temperature ($T_c$) of around 4.5K in $Bi_4O_4S_3$ [1, 2] with tetragonal $I4/mmm$ structure has generated tremendous interest among condensed matter scientific community. The novel $BiS_2$ based superconductor consists of a layered crystal structure built from $BiS_2$ layers and $Bi_4O_4(SO_4)_{1-x}$ block layers, where x indicates the deficiency of $SO_4^{2-}$ ions at interlayer sites [3]. Within few weeks, several other $BiS_2$-based superconductors, with composition $ReO_{1-x}F_xBiS_2$ (Re = La, Ce, Pr, Nd) [4-7] appeared, with $T_c$ as high as 10 K. Evidently, these compounds have a common $BiS_2$ layer, which serves as the basic building blocks of this new superconducting family. The carrier doping mechanism in the new $BiS_2$ based superconductors is seemingly the same as for the high $T_c$ cuprates and Fe-Pnictides [4-10]. There are some doubts about the exact crystal structure and superconducting composition of the $Bi_4O_4S_3$ superconductor [8]. A recent extensive study of these compounds showed that $Bi_4O_4S_3$ is actually a two-phased material [9]. The two phases are namely $Bi_2OS_2$ and $Bi_3O_2S_3$ and most likely the later one exhibits the superconductivity. Theoretical studies indicated that the Fermi level lies within the conduction bands, which are originating from the Bi 6p orbitals in the $BiS_2$ layer [11-13]. Also exotic multi-band behavior has been reported in the Hall coefficient measurements and it is shown that the charge carriers are mainly electron-type [14]. The chemical doping in the main phase of these superconducting compounds could be a powerful way to increase their $T_c$ and to understand possible mechanism of superconductivity. In one report, Ag has been doped in $Bi_4O_4S_3$ compound and it is shown that the lattice shrinks gradually with Ag doping and the $T_c$ is suppressed [15].

Keeping in view that there exists only scant reports [15] on the on-site substitution in $Bi_4O_4S_3$, we report here the effect of Se substitution at S site in bulk polycrystalline $Bi_4O_4S_{3-x}Se_x$ (x=0, 0.03, 0.09 and 0.15). All the studied samples are crystallized in near single phase and the superconductivity is seen to decrease with increase in Se content. For higher (x > 0.15) content of Se in $Bi_4O_4S_{3-x}Se_x$, the resultant material becomes clearly multi phase and the on-site substitution was not possible. It is shown that, Se doping at S site has detrimental effect on superconductivity of $Bi_4O_4S_3$.

Experimental

Polycrystalline $Bi_4O_4S_{3-x}Se_x$ (x = 0, 0.03, 0.09 and 0.15) samples are synthesized by standard solid state reaction route via vacuum encapsulation. High purity Bi, S, $Bi_2O_3$, and Se are weighed in stoichiometric ratio and ground thoroughly in a glove box under high purity argon atmosphere. The mixed powders are subsequently palletized and vacuum-sealed ($10^{-3}$ Torr) in a quartz tube. Sealed quartz ampoules are placed in Box furnace and heat treated at $400^0C$ for 10h with the typical heating rate of $2^oC$/minute, and subsequently cooled down slowly over a span of six hours to room temperature. This process was repeated twice. X-ray diffraction (*XRD*) was performed at room temperature in the scattering angular ($2\theta$) range of $10^o$-$60^o$ in equal *2θ* step of $0.02^o$ using *Rigaku Diffractometer* with *Cu $K_\alpha$* ($\lambda$ = 1.54Å). Rietveld analysis was performed using the standard *FullProf* program. The electrical transport and magnetization measurements were performed on Physical Property Measurements System (*PPMS*-14T, *Quantum Design*) as a function of both temperature and applied magnetic field.

Results and discussion

Figure 1 shows the Rietveld refined XRD patterns at a room temperature for $Bi_4O_4S_{3-x}Se_x$ (x = 0, 0.03, 0.09 and 0.15) samples. Reported Wyckoff positions [2] and space group I4/mmm with the tetragonal structure are used to refine the XRD pattern. Some impurities of Bi and $Bi_2S_3$ are also seen in the main phase of the $Bi_4O_4S_3$. The content of $Bi_2S_3$ phase slightly increases with increase in Se content. The refined lattice constants of $Bi_4O_4S_3$ are a = 3.9708(2) Å and c = 41.357(1) Å, and the same slightly increases to a = 3.9717(1) Å and c = 41.419(3) Å for x = 0.15 sample. It is clear from XRD data the lattice slightly expands due to bigger ion Se doping at S site.

Temperature dependent dc magnetic susceptibility (*M-T*) data under 10 Oe magnetic field for the studied $Bi_4O_4S_{3-x}Se_x$ samples are shown in Figure 2(a). Magnetization measurements are performed in both ZFC (Zero Field Cool) and FC (Field Cool) modes. Clear diamagnetic transitions are seen at around 4.4 K for $Bi_4O_4S_3$, 4.2 K for $Bi_4O_4S_{2.97}Se_{0.03}$, 3.7 K for $Bi_4O_4S_{2.91}Se_{0.09}$, and 3.5 K for $Bi_4O_4S_{2.85}Se_{0.15}$. It is clear that though all studied samples show superconductivity, but the $T_c$ is decreasing with increasing Se doping level. Below $T_c$, the bifurcation of the FC and ZFC data is simply due to the complicated magnetic flux pinning effects. Both the FC and ZFC magnetization data confirmed the appearance of bulk superconductivity in $Bi_4O_4S_{3-x}Se_x$ samples. Figure 2(b), represents the AC magnetic susceptibility of the $Bi_4O_4S_3$ sample, confirming the bulk superconductivity at around 4.5 K. The AC magnetic susceptibility measurements have been carried out at 333 Hz and varying amplitude of 5 - 13 Oe. It can be seen from Fig. 2b, that the imaginary part peak height is increased along with increased diamagnetism in real part of AC susceptibility with change in amplitude from 5-13 Oe. The interesting part is that with increasing AC amplitude the imaginary part peak position temperature (4.5 K) is not changed. This indicates, that the superconducting grains are well coupled and hence nearly no grain boundary contribution in the superconductivity of $Bi_4O_4S_3$. Inset of the Fig. 2b depicts the Real part of AC susceptibility being measured at applied DC bias fields of up to 4 kOe at temperature 2 K. The upper critical field ($H_{c2}$) being determined from in-field AC susceptibility measurements, is 0.5 kOe and 1.973 kOe with 50% and 90% diamagnetism criteria, respectively.

The Resistivity versus temperature ($\rho$-$T$) plots for $Bi_4O_4S_{3-x}Se_x$ (x = 0, 0.03, 0.09 and 0.15) samples in zero magnetic field are shown in Figure 3. It can be seen from the $\rho$-$T$ plots that resistivity increases slightly with increasing of Se doping. All samples showed superconductivity, with $T_c^{onset}$ of 4.5 K for $Bi_4O_4S_3$ and 4.0 K for the x = 0.15. This indicates that the superconductivity is suppressed with the Se doping in the $Bi_4O_4S_{3-x}Se_x$ compounds. It is clear indication that the superconductivity in $Bi_4O_4S_3$ compound is not due to Bi impurity. This is because the Bi content is fixed for all samples and the substitution is done only at S site. Inset of the Figure 3 shows the temperature dependent resistivity in expanded temperature range of 300 - 2.2 K. It can be seen that the all samples exhibit metallic behavior with slightly increasing resistivity for Se doped compounds.

Figure 4 (a), (b), and (c) shows the temperature dependence of resistivity under applied magnetic field for $Bi_4O_4S_{3-x}Se_x$ (x = 0, 0.03 and 0.15) samples. Superconductivity is seen at below 4.5K for pristine $Bi_4O_4S_3$ and the same decreases with Se doping. With increasing applied magnetic field, though both the $T_c^{onset}$ and $T_c(\rho = 0)$ are gradually shifted to lower temperatures, but the shift is comparatively more for the later. This results in broadening of the superconducting transition under applied field, which is similar to the high-$T_c$ layered cuprates and Fe-pnictide superconductors [16, 17]. Figure 5 exhibits the temperature dependent of upper critical field ($H_{c2}$) for $Bi_4O_4S_{3-x}Se_x$ (x = 0, 0.03 and 0.15) samples. The $H_{c2}$ is evaluated by using resistivity drops to 90% of its normal state value i.e., $\rho_n(T,H)$ at above $T_c^{onset}$. The conventional one-band Werthamer–Helfand–Hohenberg (WHH) equation $H_{c2}(0) = -0.693T_c(dH_{c2}/dT)_{T=T_c}$ has been used to evaluate $H_{c2}$. Thus calculated upper critical field $H_{c2}^{WHH}(0)$ is 2.3 Tesla for $Bi_4O_4S_3$, 2.1 Tesla for $Bi_4O_4S_{2.97}Se_{0.03}$, and 1.8 Tesla for $Bi_4O_4S_{2.85}Se_{0.15}$. It is clear that the $H_{c2}$ is decreasing with increase in Se doping concentration. The temperature derivative of resistivity ($d\rho/dT$) for the three superconducting samples $Bi_4O_4S_{3-x}Se_x$ (x = 0, 0.03 and 0.15) are shown in Fig. 6 (a), (b) and (c) respectively. Though $d\rho/dT$ peak is broadened with field yet only a single transition peak is observed, suggesting better grains coupling in these systems. The broadening of the resistive transition under applied magnetic field can be seen in terms of dissipation energy caused by the motion of vortices in the mixed state. This suggests that superconducting onset is relatively less affected than the $T_c(\rho = 0)$ state. The transition broadening is caused by the creep of vortices, which are thermally activated. The temperature dependence of resistivity in this region is governed by Arrhenius equation [18],

$$\rho(T,B) = \rho_0 \exp[-U_0/k_BT] \qquad (1)$$

Here, $\rho_0$ is the field independent pre-exponential factor, $k_B$ is the Boltzmann's constant and $U_0$ is thermally activated flux flow (TAFF) activation energy. Activation energy can be obtained from the fitted low temperature resistivity data with the Arrhenius equation. Figs. 7(a), (b) and (c) show the experimental data in terms of $\ln(\rho)$ vs. $T^{-1}$ along with the fitting to equation 1. The best fitted data give values of the activation energies to be 18.60 meV for $Bi_4O_4S_3$, 13.69 meV for $Bi_4O_4S_{2.97}Se_{0.03}$ and 13.16 meV for $Bi_4O_4S_{2.85}Se_{0.15}$ samples in applied magnetic field of 200 Oe. In low field region, though the activation energy shows weak dependence i.e. $U_0/k_B \sim H^{-0.35}$, but is strongly dependent in high field range as $U_0/k_B \sim H^{-0.85}$, it could be clearly seen that the field dependence of $U_0$ is different for lower and higher field ranges. This has been observed earlier for some other superconductors as well [16, 17]. In summery we have synthesized near single phase bulk polycrystalline samples $Bi_4O_4S_{3-x}Se_x$ (x = 0, 0.03, 0.09 and 0.15) compounds and it is found that the S site Se doping deteriorates superconductivity in $Bi_4O_4S_3$ system.

## Acknowledgements

The authors would like to thank encouragement and support from Director NPL Prof. R.C. Budhani for the superconducting materials work. One of the authors Rajveer Jha acknowledges to the CSIR for providing SRF scholarship to pursue his Ph.D. This work is also financially supported by DAE-SRC outstanding investigator award scheme on search for new superconductors.

## Figure Captions

**Figure 1:** Reitveld fitted room temperature XRD patterns of $Bi_4O_4S_{3-x}Se_x$ (x = 0, 0.03, 0.09 and 0.15) samples.

**Figure 2:** (a) Temperature dependent *dc* magnetic susceptibility (both *ZFC* and *FC)* in the applied magnetic field, H= 10 Oe plots for $Bi_4O_4S_{3-x}Se_x$ (x = 0, 0.03, 0.09 and 0.15) samples. (b) AC magnetic susceptibility in real (*M'*) and imaginary(*M"*) situations in varying amplitudes of 5–13 Oe at fixed frequency of 333 Hz for $Bi_4O_4S_3$ sample, inset shows isothermal magnetization *(MH)* for real part of AC susceptibility (*M'*) with applied field up to 4 kOe at 2 K for $Bi_4O_4S_3$ sample the upper critical field *($H_{c2}$)* is marked.

**Figure 3:** Resistivity versus temperature (*ρ* Vs *T*) plots for $Bi_4O_4S_{3-x}Se_x$ (x = 0, 0.03, 0.09 and 0.15) samples in the temperature range 2.2-5 K. Inset of the fig show the same in 2.2-300 K temperature range.

**Figure 4:** (a), (b) & (c) Temperature dependence of the resistivity *ρ(T)* under magnetic fields for the samples $Bi_4O_4S_{3-x}Se_x$ (x = 0, 0.03 and 0.15) samples respectively.

**Figure 5:** The upper critical field $H_{c2}$ found from 90% of the normalized resistivity $ρ_n(T)$ for the samples $Bi_4O_4S_{3-x}Se_x$ (x = 0, 0.03 and 0.15).

**Figure 6:** (a), (b) and (c) Temperature derivative of normalized resistivity Vs *T* of $Bi_4O_4S_{3-x}Se_x$ (x = 0, 0.03 and 0.15) samples.

**Figure 7:** Fitted Arrhenius plots of resistivity for $Bi_4O_4S_{3-x}Se_x$ (x = 0, 0.03 and 0.15) samples.

Figure 1

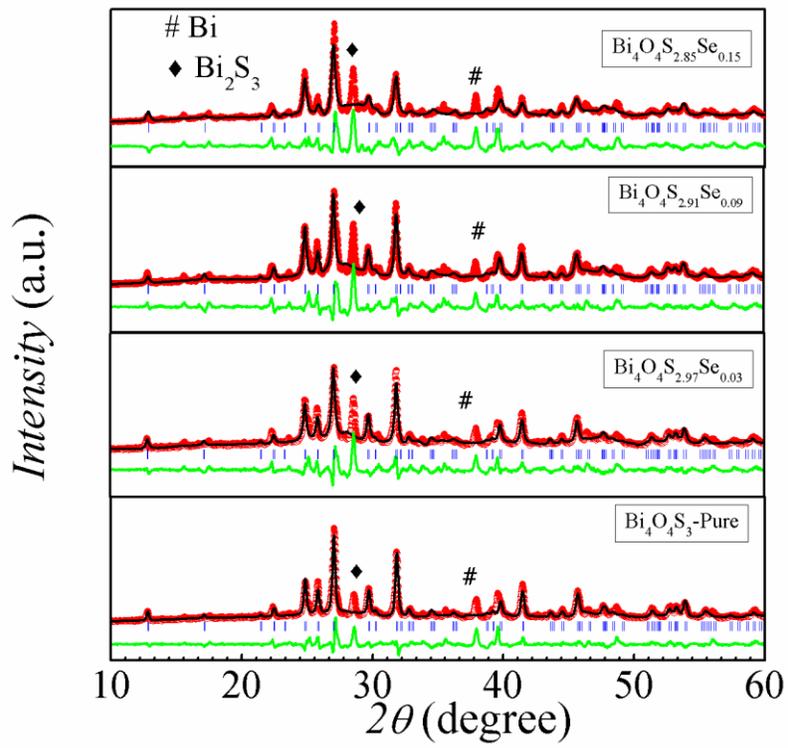

Figure 2(a)

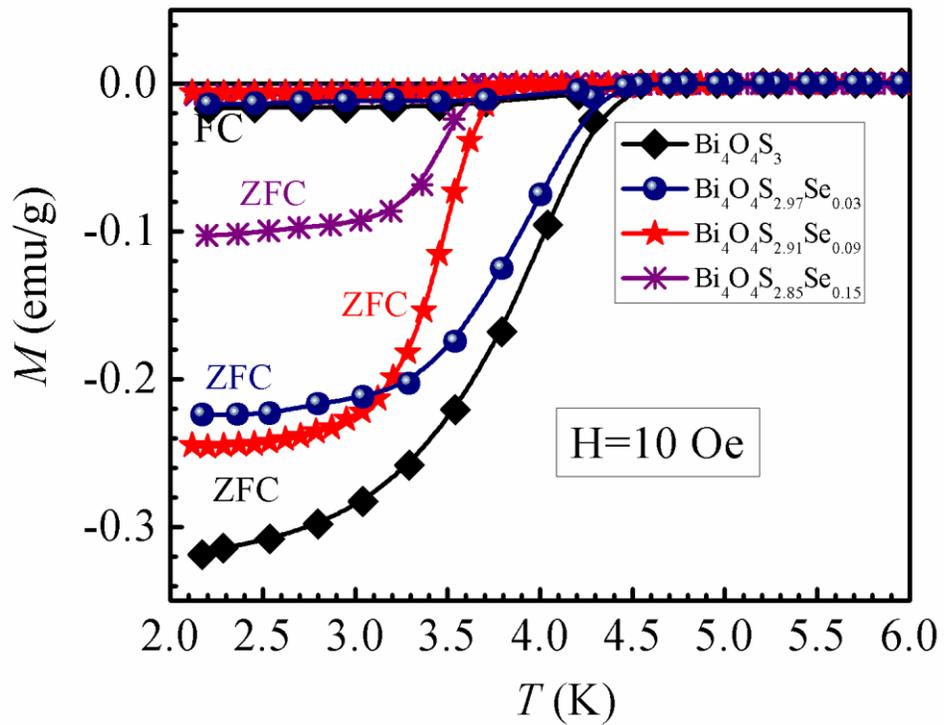

Figure 2(b)

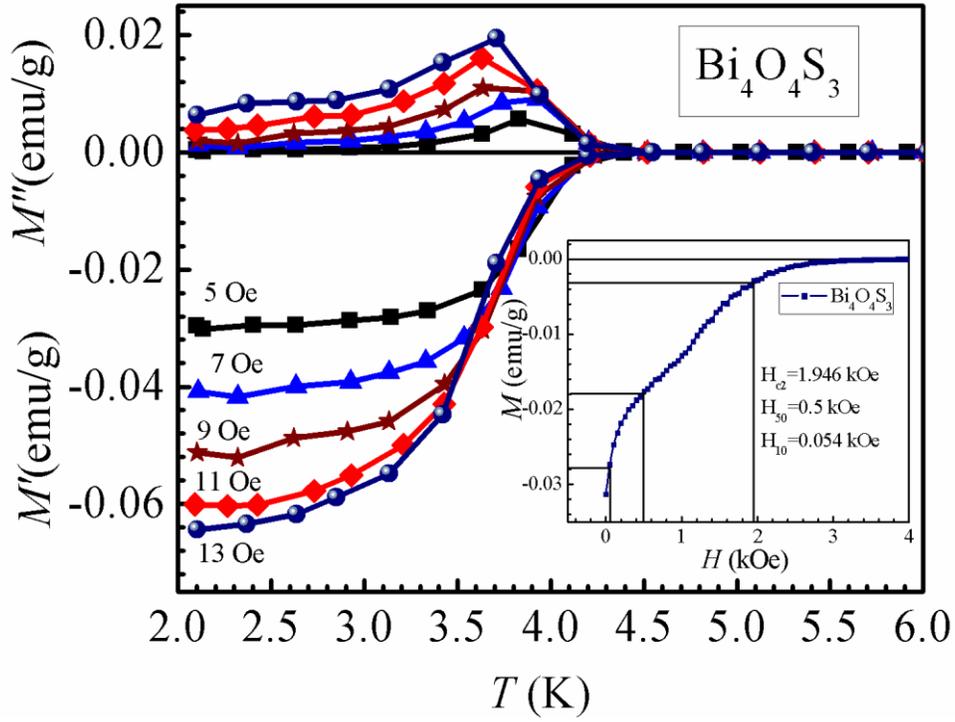

Figure 3

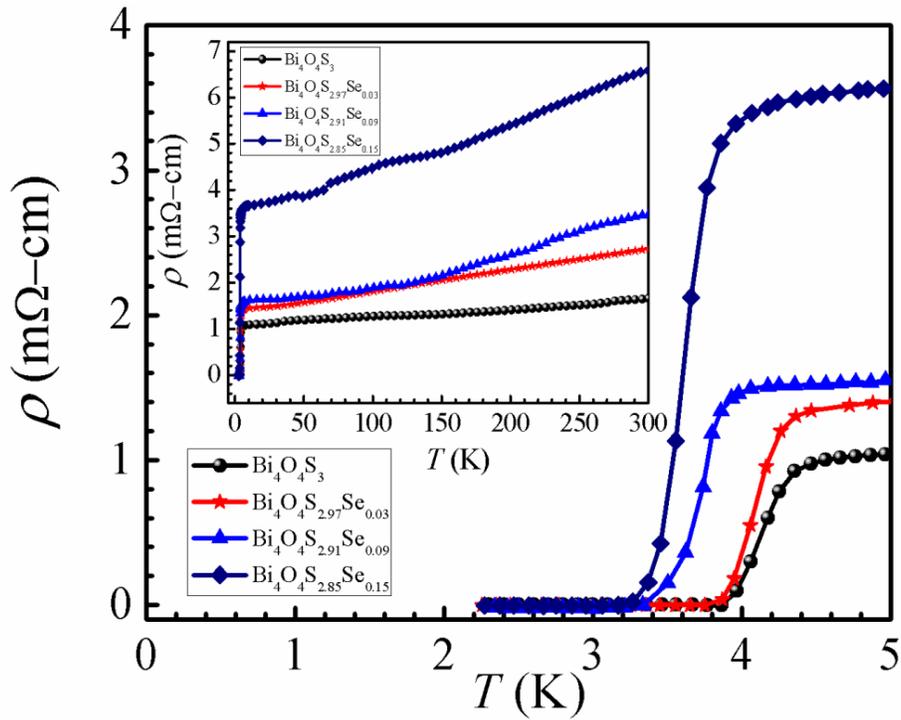

Figure 4

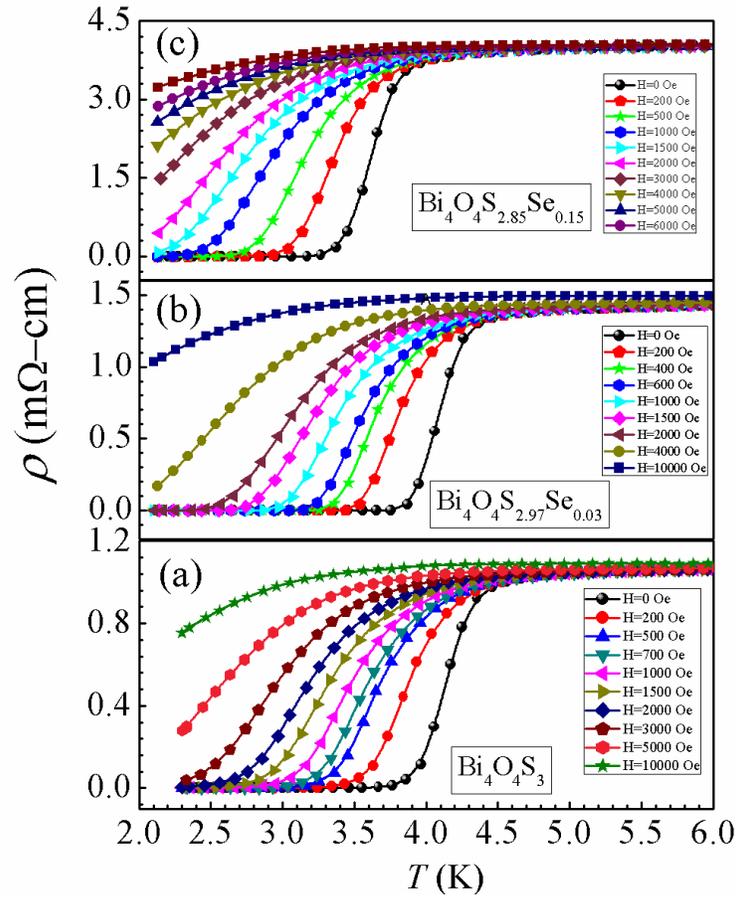

Figure 5

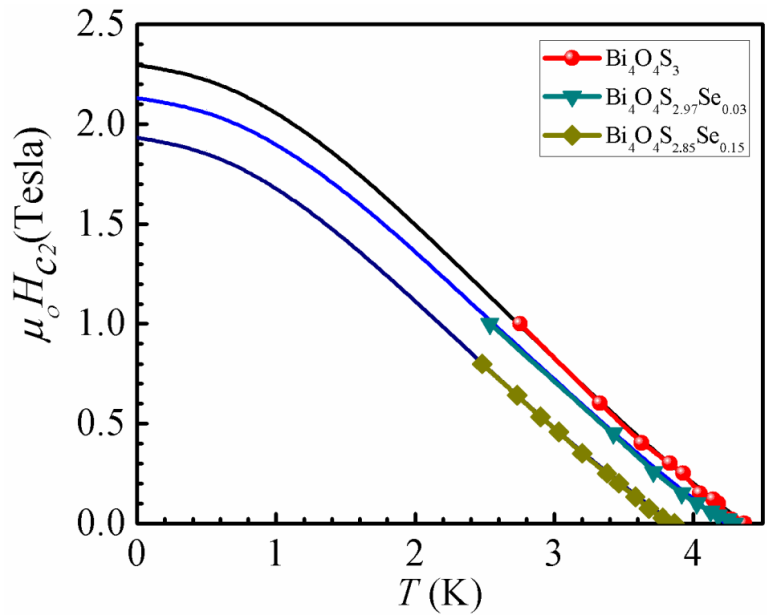

Figure 6

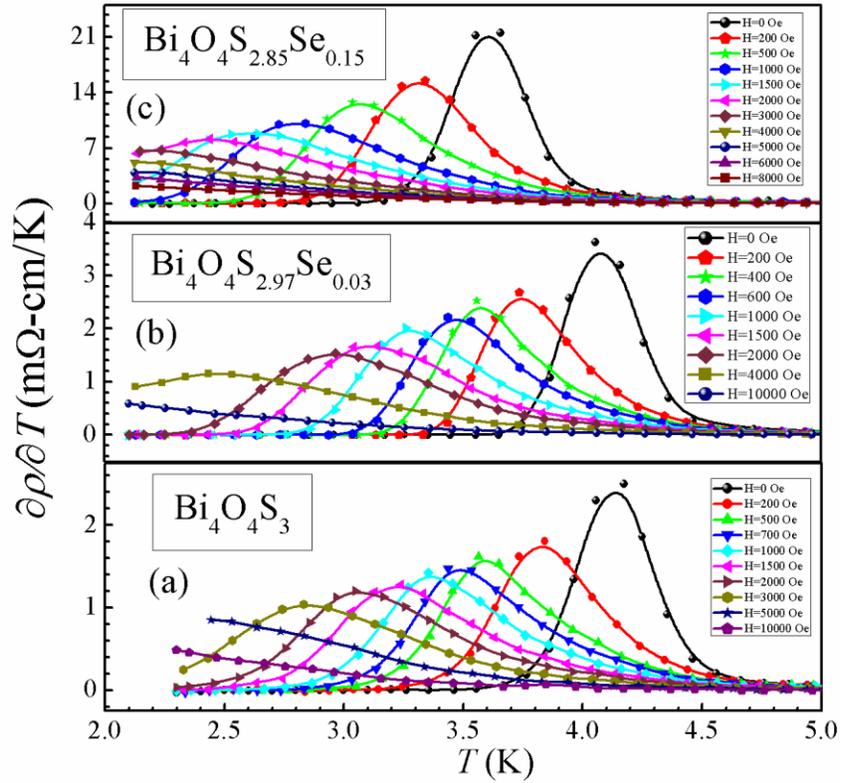

Figure 7

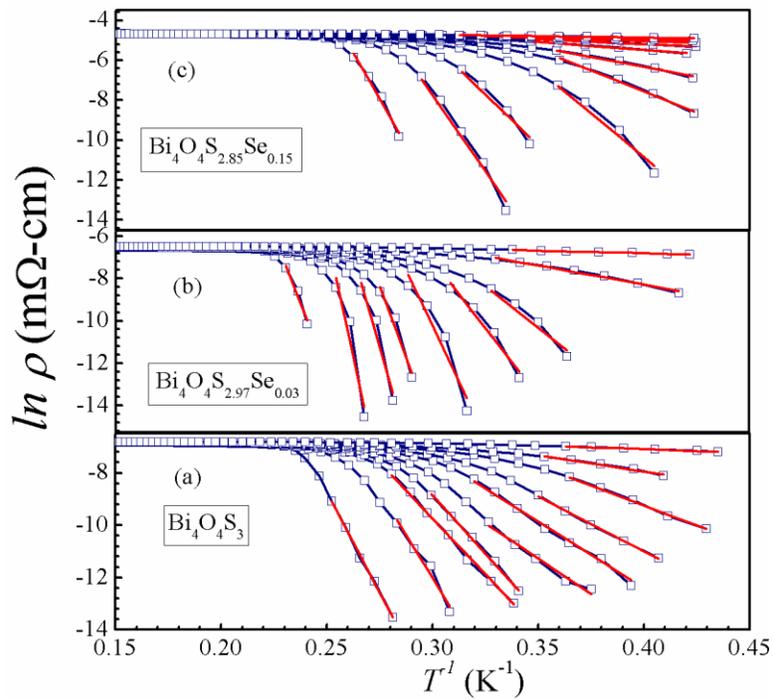